# Distortional Lifshitz Vectors and Helicity in Nematic Free Energy Density

Amelia Carolina Sparavigna[1]

[1] Department of Applied Science and Technology, Politecnico di Torino, Torino, Italy

**Abstract**: Here we discuss the free energy of nematic liquid crystals using two vectors and the helicity, with the aim of having a compact form of its density. The two vectors are due to the splay and bend distortions of the director field. They have a polar nature, whereas the helicity is a pseudoscalar.

### 1. Introduction

The free energy density of the bulk of nematics is a well-known object, fundamental for continuous theories of liquid crystals [1]. Here we propose an elementary approach to the free energy density mainly based on the role of scalars and pseudoscalars. Let us remember that pseudoscalars have odd parity upon spatial inversion, whereas scalars have even parity.

We will see that we can use two vectors, coming from the splay and bend distortion of the director field, vectors that we can define as distortional Lifshitz vectors. It is better to use the adjective "distortional" because Lifshitz vectors are specific vectors of the relativistic gravitational field. These distortional vectors have a polar nature. In addition them we need a pseudoscalar, the helicity, to write the nematic free energy density. The helicity squared gives rise to the twist contribution. Note that the distortional vectors that we will use do not depend on a polar or axial nature of the director field of nematic liquid crystals. Even if this choice does not influence the final result, we will see that the director behaves like a vector potential.

### 2. The free energy of nematics

Let as consider the bulk free energy as discussed in the book by Landau and Lifshitz [2]. The free energy of a nematic liquid crystal can have only scalar terms of the director field $\vec{n}$ and its derivatives. The true scalars can be obtained from the following product of derivatives:

$$\frac{\partial n_k}{\partial x_i}\frac{\partial n_l}{\partial x_m} \quad (1)$$

where $x_i$ are the coordinates of the frame. Moreover:

$$\frac{\partial}{\partial x_i}\vec{n}^2 = 2n_k\frac{\partial n_k}{\partial x_i} = 0 \quad (2)$$

because $\vec{n}^2 = 1$. After contraction of indices or multiplying by $\vec{n}$, we find the invariants:

$$[(\vec{n}\cdot\nabla)\vec{n}]^2 \quad ; \quad \frac{\partial n_k}{\partial x_i}\frac{\partial n_k}{\partial x_i} \quad ; \quad (\nabla\cdot\vec{n})^2 \quad ; \quad \frac{\partial n_k}{\partial x_i}\frac{\partial n_i}{\partial x_k} \quad (3)$$

The last two terms differ by a divergence. Since the divergence goes into a surface term, we do not consider among the bulk terms. Therefore we can keep only the divergence squared $(\nabla\cdot\vec{n})^2$. The second term is the sum of the divergence squared and $(\vec{n}\cdot\nabla\times\vec{n})^2$. Another possible contribution is $(\vec{n}\cdot\nabla\times\vec{n})(\nabla\cdot\vec{n})$, and we have also that $[(\vec{n}\cdot\nabla)\vec{n}]^2 = [\vec{n}\times\nabla\times\vec{n}]^2$. Therefore the free energy density is given by Landau and Lifshitz as:

$$F = F_o + b\vec{n}\cdot\nabla\times\vec{n} + \frac{a_1}{2}(\nabla\cdot\vec{n})^2 + \frac{a_2}{2}(\vec{n}\cdot\nabla\times\vec{n})^2 + \frac{a_3}{2}[\vec{n}\times\nabla\times\vec{n}]^2 + a_{12}(\vec{n}\cdot\nabla\times\vec{n})(\nabla\cdot\vec{n}) \quad (4)$$

The first term is a contribution that does not depend on the director field. As [2] is telling, $(\vec{n}\cdot\nabla\times\vec{n})$ is

Amelia Carolina Sparavigna (Correspondence) ✉
amelia.sparavigna@polito.it
DOI: 10.18483/ijSci.211

a pseudovector. The second term in (4) must have a coefficient *b* which is a pseudoscalar. As a consequence, this term exists in cholesteric nematic. The last tem is equal to zero, because in nematics we require the symmetry $\vec{n} \to -\vec{n}$. The nematic bulk free energy density is:

$$F = F_o + \frac{a_1}{2}(\nabla \cdot \vec{n})^2 + \frac{a_2}{2}(\vec{n} \cdot \nabla \times \vec{n})^2 + \frac{a_3}{2}[\vec{n} \times \nabla \times \vec{n}]^2 \qquad (5)$$

In his book [3], S.A. Pikin is starting the discussion of free energy density from the order parameter tensorial field, which is given by:

$$Q_{ik}(\vec{r}) = Q(\vec{r})\left(n_i(\vec{r})n_k(\vec{r}) - \frac{1}{3}\delta_{ik}\right) \qquad (6)$$

In the isotropic phase $Q = 0$. The orientation order parameter $Q$, which is a function of temperature, far from phase transition points is not subjected to strong thermal fluctuations. For this reason, $Q$ can be viewed as a constant which is characterizing the features of the anisotropic medium at a fixed temperature. At the same time, the director $\vec{n}$ is subjected to appreciable thermal fluctuations and changes relatively easily under the action of external fields [3]. This circumstance is one and the most important reason for the instabilities of the orientational structures, appearing in the liquid crystal both as a result of temperature changes and under external action.

Under the condition that function $\vec{n}(\vec{r})$ is slowly varying along the bulk, i.e., the derivatives of this function with respect to the coordinates are small, the orientational free energy density *F* of the deformed nematic can be expanded in powers of the derivatives of function $\vec{n}(\vec{r})$. Such an expansion of a scalar quantity must contain scalar combinations of the vector $\vec{n}(\vec{r})$ and its derivatives. As previously told, $\vec{n}$ is a unit vector, which can enter only in even combinations due to the equivalence $\vec{n} \to -\vec{n}$. In addition, the total derivatives that make a contribution to the surface energy and not to the volume energy of the body must be dropped in the expression of *F*.

$\nabla \cdot \vec{n}$ is odd in $\vec{n}$, moreover it is a surface term, and $\vec{n} \cdot \nabla \times \vec{n}$ is a pseudoscalar. We use these terms as squared, $(\nabla \cdot \vec{n})^2$ and $(\vec{n} \cdot \nabla \times \vec{n})^2$, as well as the scalar products of the vectors formed by the vector $\vec{n}$ and its first derivatives. There are three such independent vectors:

$$\nabla \times \vec{n} \quad ; \quad \vec{n}\nabla \cdot \vec{n} \quad ; \quad [\vec{n} \times \nabla \times \vec{n}] = -(\vec{n} \cdot \nabla)\vec{n} \qquad (7)$$

Additional scalar invariants are:

$$(\nabla \times \vec{n})^2 \quad ; \quad (\vec{n} \cdot \nabla \times \vec{n})\nabla \cdot \vec{n} \quad ; \quad [\vec{n} \times \nabla \times \vec{n}]^2 \quad ; \quad (\vec{n} \cdot \nabla \times \vec{n})^2 \qquad (8)$$

Since the combination $(\vec{n} \cdot \nabla \times \vec{n})\nabla \cdot \vec{n}$ is odd with respect to $\vec{n}$ and a pseudoscalar, we do not use it. Moreover, as a result of the relation (see Appendix A):

$$(\nabla \times \vec{n})^2 = (\vec{n} \cdot \nabla \times \vec{n})^2 + (\vec{n} \times \nabla \times \vec{n})^2 \qquad (9)$$

We have again:

$$F = F_o + \frac{a_1}{2}(\nabla \cdot \vec{n})^2 + \frac{a_2}{2}(\vec{n} \cdot \nabla \times \vec{n})^2 + \frac{a_3}{2}[\vec{n} \times \nabla \times \vec{n}]^2 \qquad (10)$$

It is a combination of these three terms that can be used to represent an arbitrary deformation in a nematic liquid crystal. The coefficients of this combination are the elastic constant of splay, twist and bend respectively. It is often the case that all these three constants are of the same order of magnitude: the combination is commonly approximated to have $K = a_1 = a_2 = a_3$. This approximation is commonly referred to as the one-constant approximation and is used predominantly because the free energy density simplifies to the compact form:

$$F = F_o + \frac{K}{2}\left[(\nabla \cdot \vec{n})^2 + (\nabla \times \vec{n})^2\right] \qquad (11)$$

For computation, this form is quite easy but it is mixing the distortional contributions.

We could imagine the uniform state as energetically favored, but this state is unstable relative to spatial modulations of the director. The presence of a

$$F_{chol} = \frac{K_2}{2}(q + \vec{n} \cdot \nabla \times \vec{n})^2 \qquad (12)$$

in the free energy density. Let us note that $q$ has the dimensions of a wave number. Pikin is remarking that this expression contains the pseudoscalar $(\vec{n} \cdot \nabla \times \vec{n})$ which is giving the helical structure. In Eq.4, the term appears as $b\vec{n} \cdot \nabla \times \vec{n}$, requiring $b$ as a pseudoscalar, according to the chiral nature of molecules. In (12), we have the contribution $K_2 q \vec{n} \cdot \nabla \times \vec{n}$: since $K_2$ is a scalar, this means that $q$ is a pseudoscalar (the microscopic origin of the chiral term had been discussed in Ref.4).

As we have seen, the free energy density is assumed as a true scalar. Or, as it is told in [5], the contributions to the free energy density "of course, have to be true invariant scalar". Let me open a discussion on the densities, as proposed by Dalton Schnack, in his paper published in the Lectures Notes in Physics [6]. In this paper, it is told that since the volume is a scalar triple product, that is, the scalar product of a vector by a surface vector, according to a parity transformation it is more properly described as a pseudoscalar. As a consequence of the pseudoscalar nature of the volume, the mass density and the pressure, the ratio of the internal energy and volume, behave like pseudoscalars. That is, these quantities have an odd parity. The reference is concluding telling that the resulting expressions that we are using must have a consistent parity. The volume discussed in [6] is of course the oriented volume, which averages out to zero. However, we can use the modulus of the oriented volume and therefore assume the energy density like a scalar.

After this general review of the bulk free energy density, let us try to obtain its terms using two vectors that we can define as the distortional Lifshitz vectors, and the helicity. In fact, the distortional vectors are contained in some Lifshitz invariants. Therefore, before discussing these vectors, let us shortly review the Lifshitz invariants and their role in the appearance on modulated structures.

### 3. Lifshitz invariants and modulated structures
According to Lifshitz [7], near a phase transition point, the system may be unstable with respect to distortions of the appropriate order parameter. This macroscopic inhomogeneity of the director in the cholesteric liquid crystals at distances that are large compared to molecular dimensions is related with the existence of an invariant of the form:

instability may develop if the irreducible representation allows a quadratic anti-symmetric combination, linear both in the order parameter components and in their gradients. Therefore the contributions to the free-energy density of terms in the derivatives of the order parameter are governing the appearance of spatially modulated structure in magnetic materials and liquid crystals. And in fact, it is possible to see the same invariant in the free energy, the Lifshitz invariant, as the responsible of undulated patterns.

A specific formulation by vectors of the Lifshitz invariants was first proposed for some magnetic structures characterized by a modulation of the spin arrangements [8,9]. Within a continuum approximation of magnetic properties, the interactions responsible for these modulations are expressed by inhomogeneous invariants. These contributions to the free magnetic energy, involving first derivatives of magnetization with respect to spatial coordinates, are defined as the inhomogeneous Dzyaloshinskii-Moriya interactions [10,11]. These interactions are linear with respect to the first spatial derivatives of the magnetization $\vec{M}$ in an anti-symmetric mathematical form.

The structure of the Lifshitz invariant is, in the case of the inhomogeneous Dzyaloshinskii-Moriya interaction, a product of three vectors. The three vectors are: a vector $\vec{D}$ representing an internal or external field, a vector $\vec{M}$ representing the local order parameter, and $\nabla$ operating on the order parameter components. The product has the following form:

$$F_L = \vec{D} \cdot \left[\vec{M}(\nabla \cdot \vec{M}) - (\vec{M} \cdot \nabla)\vec{M}\right] \qquad (13)$$

We used the Dzyaloshinskii-Moriya interactions in 1996, to study the field-induced phase transition of $BiFeO_3$ [12]. An antiferromagnetic vector $\vec{L}$ characterizes the $BiFeO_3$ spin structure. A Landau-Ginzburg energy density of the spin structure was given in Ref.12, using the following vector:

$$\vec{A} = \vec{L}(\nabla \cdot \vec{L}) - (\vec{L} \cdot \nabla)\vec{L} \qquad (14)$$

One term of the free energy density is the Lifshitz invariant given as:

$$F_L = -\overline{\alpha}\vec{P}_S \cdot \vec{A} \qquad (15)$$

In (15), $\overline{\alpha}$ is the inhomogeneous relativistic constant and $\vec{P}_S$ the spontaneous polarization.

In Ref.12, we investigated the influence of an electric field on the spatially modulated spin structure (SDW state), using the analogy with nematic liquid crystals to study magnetic materials.

### 4. The flexoelectric vectors in nematics

Let us consider a nematic liquid crystal and assume the order parameter described by the director field $\vec{n}$, which is giving the local mean orientation of the molecules. Vector $\vec{A}$ in Eq.14 can then be written in the following form:

$$\vec{A} = \vec{n}(\nabla \cdot \vec{n}) - (\vec{n} \cdot \nabla)\vec{n} = \vec{n}\,\mathrm{div}\,\vec{n} + \vec{n} \times \mathrm{rot}\,\vec{n} \qquad (16)$$

This vector is well known in the physics of liquid crystals. It is encountered in the structure of the flexoelectric contribution to the free energy which is $f_{flexo} = -\vec{P} \cdot \vec{E}$. Flexoelectricity is a property of liquid crystals similar to the piezoelectric effect [13]. A distortion of the director field can induce a macroscopic polarization within the material [11,13]. The polarization vector $\vec{P}$, a polar vector, in the flexoelectric term is then described with a distortion in the nematic director field:

$$\vec{P} = e_S\,\vec{n}(\nabla \cdot \vec{n}) - e_B(\vec{n} \cdot \nabla)\vec{n} = e_S\,\vec{n}\,\mathrm{div}\,\vec{n} + e_B\,\vec{n} \times \mathrm{rot}\,\vec{n} \qquad (17)$$

The two terms in the polarization vector are due to the splay and the bend contribution.

The coupling of the flexoelectric polarization $\vec{P}$ with an external electric field results in the appearance of a periodic distortion, as the term shown by Eq.(15) in the free energy density of BiFeO$_3$, where we have the coupling of vector $\vec{A}$ with a spontaneous polarization.

In the flexoelectricity, the polarization is induced by a not uniform deformation. If we consider the flexoelectric coupling, the electric field $\vec{E}$ is a polar

$$\vec{A} = \vec{n}\,\mathrm{div}\,\vec{n} + \vec{n} \times \mathrm{rot}\,\vec{n} = \vec{n}\,\nabla \cdot \vec{n} + \vec{n} \times \nabla \times \vec{n}$$

Let us note that assuming $\vec{n}$ a vector, $\vec{A}$ is a vector. Behaving $\vec{n}$ like a pseudovector $\vec{A}$ is a vector. Let us consider the first term in $\vec{A}$: the divergence of a pseudovector is a pseudoscalar, which multiplied by the pseudovector, gives a vector (the gradient operator transforms like a polar vector). The second term is the cross product of a pseudovector and a vector, therefore it is a vector. The fact that the Lifshitz vector is a true vector is in agreement with its appearance in the polarization of the flexoelectric effect, which is coupled with the electric field. The scalar product of the electric field, a true vector, and vector, the polarization is a polar vector, and then the additional term in the free energy density is a scalar. $\vec{A}$ has the nature of a vector, independently of the choice of $\vec{n}$ as a vector or a pseudovector.

### 5. The distortional Lifshitz vectors and the free energy density

As we have previously discussed, the Lifshitz invariant can be defined using the vector:

$$\qquad (18)$$

$\vec{A}$ vector gives a scalar term for the bulk free energy density.

Therefore, let us consider some scalars that we can have after this vector:

$$\vec{A} \cdot \vec{A} = (\vec{n}\,\nabla \cdot \vec{n})^2 + (\vec{n} \times \nabla \times \vec{n})^2 \qquad (19)$$

because $\vec{n} \cdot (\vec{n} \times \nabla \times \vec{n})$ is equal to zero for the properties of the cross product. The two terms are the splay and the bend contributions to the free energy. Let us try:

$$\vec{n} \cdot \vec{A} = \vec{n} \cdot (\vec{n}\,\nabla \cdot \vec{n} + \vec{n} \times \nabla \times \vec{n}) = \vec{n} \cdot \vec{n}\,\nabla \cdot \vec{n} + \vec{n} \cdot (\vec{n} \times \nabla \times \vec{n}) = \nabla \cdot \vec{n} \qquad (20)$$

This term does not appears because it does not satisfy symmetry $\vec{n} \to -\vec{n}$. The term squared is a correction of the free energy density.
Let us consider:

$$\vec{n} \times \vec{A} = \vec{n} \times (\vec{n}\,\nabla \cdot \vec{n} + \vec{n} \times \nabla \times \vec{n}) = \vec{n} \times \vec{n}\,\nabla \cdot \vec{n} + \vec{n} \times (\vec{n} \times \nabla \times \vec{n}) = \vec{n} \times (\vec{n} \times \nabla \times \vec{n}) \qquad (21)$$

Again this term does not appears because of the required symmetry $\vec{n} \to -\vec{n}$. The term squared:

$$(\vec{n} \times \vec{A})^2 = (\vec{n}(\vec{n} \cdot \nabla \times \vec{n}) - \nabla \times \vec{n}(\vec{n} \cdot \vec{n}))^2 = (\vec{n}(\vec{n} \cdot \nabla \times \vec{n}) - \nabla \times \vec{n})^2 =$$
$$= (\vec{n} \cdot \nabla \times \vec{n})^2 - 2\vec{n}(\vec{n} \cdot \nabla \times \vec{n}) \cdot \nabla \times \vec{n} + (\nabla \times \vec{n})^2 = -(\vec{n} \cdot \nabla \times \vec{n})^2 + (\nabla \times \vec{n})^2 \quad (22)$$

We have that: $(\nabla \times \vec{n})^2 = (\vec{n} \cdot \nabla \times \vec{n})^2 + (\vec{n} \times \nabla \times \vec{n})^2$ (see Appendix A), and then:

$$(\vec{n} \times \vec{A})^2 = -(\vec{n} \cdot \nabla \times \vec{n})^2 + (\nabla \times \vec{n})^2 = -(\vec{n} \cdot \nabla \times \vec{n})^2 + (\vec{n} \cdot \nabla \times \vec{n})^2 + (\vec{n} \times \nabla \times \vec{n})^2 = (\vec{n} \times \nabla \times \vec{n})^2$$
(23)

From the previous equation, we see that we can obtain only two terms of the free energy density:

$$F = F_o + \frac{a_1}{2}(\nabla \cdot \vec{n})^2 + \frac{a_2}{2}(\vec{n} \cdot \nabla \times \vec{n})^2 + \frac{a_3}{2}[\vec{n} \times \nabla \times \vec{n}]^2 \quad (24)$$

Splitting $\vec{A}$ vector in two vectors:

$$\vec{A} = \vec{A}_1 + \vec{A}_3 \quad ; \quad \vec{A}_1 = \vec{n}\, \nabla \cdot \vec{n} \quad ; \quad \vec{A}_3 = \vec{n} \times \nabla \times \vec{n} \quad (25)$$

We see that $\vec{A}_1 \cdot \vec{A}_3 = 0$ and $\vec{A} \cdot \vec{A} = \vec{A}_1^2 + \vec{A}_3^2$. These are the two distortional Lifshitz vectors, which are polar orthogonal vectors, able to provide two terms of the free energy (splay and bend), but not the twist term $(\vec{n} \cdot \nabla \times \vec{n})^2$. In fact, this is a helicity term, which has a pseudoscalar nature, squared to be consistent with the true scalar nature of the free energy density.

Let us therefore define $\vec{A}_1, \vec{A}_3$ as the distortion vectors in general and add to them a pseudoscalar to obtain all the contributions of the bulk free energy.

**6. The helicity density**
In studying fluids, the helicity density is defined as:

$$h = \vec{v} \cdot \nabla \times \vec{v} \quad (26)$$

where $\vec{v}$ is the velocity. The helicity density is a pseudoscalar, having the same form of $(\vec{n} \cdot \nabla \times \vec{n})$, that we have squared in the free energy density. And this is the pseudoscalar, that we are not able to obtain using the Lifshitz vectors. The pseudo nature of this term is forbidding its creation from the two orthogonal Lifshitz vectors. In particle physics, the helicity is the projection of the spin onto the direction of momentum. For massless particles, such as the photon, the particle appears to spin in the same direction along its axis of motion regardless of point of view of the observer. That is, the helicity is a relativistic invariant.
In electromagnetism, the magnetic helicity density is given by:

$$h = \vec{A} \cdot \nabla \times \vec{A} \quad (27)$$

Polar vector $\vec{A}$ is the vector potential. It is a vector field and its curl is the magnetic field: $\vec{B} = \nabla \times \vec{A}$ If a vector field admits a vector potential $\vec{A}$, then the equality $\nabla \cdot (\nabla \times \vec{A}) = 0$ implies that $\vec{B}$ must be a solenoid vector field. The vector potential is not unique: if $\vec{A}$ is a vector potential, so is

$$\vec{A} + \nabla f \quad (28)$$

In (28), $f$ is any continuously differentiable scalar function. This follows from the fact that the curl of the gradient is zero. This non-uniqueness leads to a degree of freedom in the formulation of electrodynamics which implies the choice of a gauge. In the case of the nematic liquid crystals, the director $\vec{n}$ behaves like $\vec{A}$. And imposing $\vec{n}$ as a unit vector, we are fixing the gauge.
The free energy density of a nematic can be written in the following form:

$$F = F_o + \frac{a_1}{2}A_1^2 + \frac{a_3}{2}A_3^2 + \frac{a_2}{2}h^2 \quad (29)$$

Here I used a helicity different from the conserved quantity in nematic liquid crystal flows, which is given in Ref.14. In the case of a cholesteric:

$$F = F_o + bh + \frac{a_1}{2}A_1^2 + \frac{a_3}{2}A_3^2 + \frac{a_2}{2}h^2 \quad (30)$$

Equation (29) is a quite compact form of linear combinations of scalars. This form of the free energy density has been obtained using the triplet ($\vec{A}_1, \vec{A}_3, h$), of two polar vectors and a pseudoscalar. A motivation for the use of such a triplet is to increase the analogy of the distortional energy density with field theories. Second-order elasticity and surface terms are under investigation to obtain some similar results from terms already proposed in [15].

**Appendix A**

Let us show that: $(\nabla \times \vec{n})^2 = (\vec{n} \cdot \nabla \times \vec{n})^2 + (\vec{n} \times \nabla \times \vec{n})^2$. Let us use: $(\nabla \times \vec{n})_x = R_x$, $(\nabla \times \vec{n})_y = R_y$ and $(\nabla \times \vec{n})_z = R_z$. We have:

$(\vec{n} \cdot \nabla \times \vec{n})^2 + (\vec{n} \times \nabla \times \vec{n})^2 = (n_x R_x + n_y R_y + n_z R_z)^2 +$
$+ (n_y R_z - n_z R_y)^2 + (n_x R_z - n_z R_x)^2 + (n_x R_y - n_y R_x)^2 =$
$= n_x^2 R_x^2 + n_y^2 R_y^2 + n_z^2 R_z^2 + 2 n_x n_y R_x R_y + 2 n_x n_z R_x R_z + 2 n_y n_z R_y R_z +$
$+ n_y^2 R_z^2 + n_z^2 R_y^2 - 2 n_y n_z R_y R_z + n_x^2 R_z^2 + n_z^2 R_x^2 - 2 n_x n_z R_x R_z + n_x^2 R_y^2 + n_y^2 R_x^2 - 2 n_x n_y R_x R_y =$
$= n_x^2 R_x^2 + n_y^2 R_y^2 + n_z^2 R_z^2 + n_y^2 R_z^2 + n_z^2 R_y^2 + n_x^2 R_z^2 + n_z^2 R_x^2 + n_x^2 R_y^2 + n_y^2 R_x^2 =$
$= n_x^2 R_x^2 + n_y^2 R_x^2 + n_z^2 R_x^2 + n_x^2 R_y^2 + n_y^2 R_y^2 + n_z^2 R_y^2 + n_x^2 R_z^2 + n_y^2 R_z^2 + n_z^2 R_z^2 = R_x^2 + R_y^2 + R_z^2$
$= (\nabla \times \vec{n})^2$

**References**
[1] G. Barbero and L.R. Evangelista, An Elementary Course on the Continuum Theory for Nematic Liquid Crystals, World Scientific, 2001.
[2] L.D. Landau and E.M. Lifshitz, *Statistical Physics*, Pergamon Press, Oxford, 1980.
[3] S.A. Pikin, *Structural Transformations In Liquid Crystals*, Gordon and Breach Science Publishers, 1991
[4] A. B. Harris, Randall D. Kamien, T. C. Lubensky, Phys. Rev. Lett. 78, 1476–1479 (1997), Microscopic Origin of Cholesteric Pitch, http://arxiv.org/pdf/cond-mat/9607084v3.pdf
[5] Mark Warner, Eugene Michael Terentjev, Liquid Crystal Elastomers, Oxford University Press, 31/mag/2007
[6] Dalton Schnack, Review of Scalars, vectors, tensors and dyads, Lectures Notes in Physics, Springer, 780, 2009, pp.5-18.
[7] E.M. Lifschitz, On the theory of second-order phase transitions. Zh. Eksp. Teor. Fiz. 1941, 11, 255-269.
[8] I.E. Dzyaloshinskii, Theory of helicoidal structures in antiferromagnets .1. Nonmetals, Sov. Phys. JETP 1964, 19, 960.
[9] T. Moriya, Anisotropic Superexchange Interaction and Weak Ferromagnetism, Phys. Rev. 1960, 120, 91.
[10] A.N. Bogdanov, U.K. Rößler, M. Wolf, and K.-H. Müller, Magnetic structures and reorientation transitions in noncentrosymmetric uniaxial antiferromagnets, arXiv:cond-mat/0206291, 2002.
[11] A. Sparavigna, Role of Lifshitz Invariants in Liquid Crystals, Materials, 2009, 2(2), 674-698.
[12] A.Sparavigna, A.Strigazzi, A.K.Zvezdin, Phys. Rev. B 1994, 50, 2953.
[13] R.B. Meyer, Piezoelectric effects in liquid crystals, Physical Review Letters, Vol. 22(18), 1969, 918-921.
[14] François Gay-Balmaz, Cesare Tronci, The helicity and vorticity of liquid crystal flows, arXiv, 2010, http://arxiv.org/abs/1006.2984
[15] G. Barbero, A. Sparavigna, A. Strigazzi, The structure of the distortion free-energy density in nematics: second- order elasticity and surface terms, Nuovo Cimento, 12D, 1990, 1259-1272.